\documentclass[aps,pre,twocolumn,byrevtex]{revtex4-1}

 \usepackage{orcidlink}

 \usepackage{pdfpages}
 \usepackage{color}
 \usepackage{hyperref}
 \usepackage{epsfig}
 \usepackage{amsmath}
 \usepackage{amsthm}
 \usepackage{amsfonts}
 \usepackage{amssymb}
 \usepackage{wasysym}

 \usepackage[english]{babel}
 \usepackage[utf8]{inputenc}

\makeatletter
\patchcmd{\@outputpage@head}{\@ifx{\LS@rot\@undefined}{}{\LS@rot}}{}{}{}
\makeatother

\def\be{\begin{equation}}
\def\eea{\end{eqnarray}}
\def\ee{\end{equation}}
\def\bea{\begin{eqnarray}}
\def\ea{\end{array}}
\def\ba{\begin{array}}

\newcommand{\exval}[1]{\mbox{$\left\langle \, {#1}\, \right\rangle$}}

\newcommand{\bel}[1]{\begin{equation}\label{#1}}

\newcommand{\bs}{\backslash}

\newcommand{\D}{\mathrm d}

\def\zzz{{\mathchoice {\hbox{$\sf\textstyle Z\kern-0.4em Z$}}
{\hbox{$\sf\scriptstyle Z\kern-0.3em Z$}}
{\hbox{$\sf\scriptscriptstyle Z\kern-0.2em Z$}}
{\hbox{$\sf\textstyle Z\kern-0.4em Z$}}}}

\begin{document}

\title{
Mahler measures, elliptic curves and $L$-functions for the free
energy of the Ising model
}

\author{Gandhimohan M. Viswanathan\orcidlink{0000-0002-2301-5593 }}

\affiliation{Department of Physics
and \mbox{National Institute of Science and Technology of Complex Systems,}
Federal University of  Rio Grande do Norte, 59078-970
Natal--RN, Brazil}

\begin{abstract}

This work establishes links between the Ising model and
elliptic curves via Mahler measures. 
First, we reformulate the partition function of the Ising model on the
square, triangular and honeycomb lattices in terms of the Mahler measure
of a Laurent polynomial whose variety's projective closure 
defines an elliptic curve. 
  Next, we obtain hypergeometric formulas for the partition functions
  on the triangular and honeycomb lattices and review the known
  series for  the square lattice.
  Finally, at specific temperatures we express  the free energy
in terms of a  Hasse-Weil
$L$-function of an elliptic curve.
At the critical point of the phase transition on all three lattices,
we obtain the free energy more simply in terms of a Dirichlet
$L$-function.  These findings suggest that the connection between
statistical mechanics and analytic number theory may run deeper than
previously believed.

\end{abstract}

\maketitle

\section{Introduction}
In the last two decades, Mahler
measures~\cite{smyth-book,zudilin-book} have found applications across
diverse areas of physics, including  crystal melting~\cite{zahabi},
instantons~\cite{stienstra-instanton},
mirror symmetry~\cite{stienstra-instanton},
 toric quiver gauge theories~\cite{bao2022,zahabi},
extremal (BPS) black holes~\cite{zahabi},
topological string
theories~\cite{bao2022,zahabi},
networked control systems~\cite{chesi2013} and 
random walks on lattices~\cite{viswan2017-pre,zudilin-book,guttmann2012}.
In statistical mechanics, they play key roles in spanning tree
generating functions, lattice Green functions, and partition
functions of lattice
models~\cite{viswan2017-pre,guttmann2012,wu-of-glasser,glasser-of-glasser,shrock-of-glasser,glasser-cubic,stienstra-dimer}.
Independently of these advances in physics, mathematicians in the
early 1990s discovered a remarkable  connection between Mahler measures
and
\mbox{$L$-functions} of elliptic
curves~\cite{boyd1998,boyd1981,smyth-book,lalin2007,villegas-1999,zudilin-book,glasser-cubic,stienstra-dimer,stienstra-instanton}.
A natural question thus arose regarding  the possibility of a direct link
between quantities of interest in statistical mechanics on the one
hand and $L$-functions on the other.
For example, Stienstra~\cite{stienstra-dimer} observed that ``there may be some mysterious
link between the partition function of a dimer model and the
$L$-function of its spectral curve,'' and proceeded to give a partial
answer to this question by finding $L$-functions for several dimer
models.
We further investigate this ``mysterious link'' and show that it
generalizes in an unanticipated and intriguing way to the Ising model
--- arguably the most important and widely studied model in
statistical mechanics.  Indeed, the findings presented in this work
suggest that the connection extends deeper than previously known
or
hypothesized~\cite{viswan2017-pre,guttmann2012,ising-l-function,ising-sigma,ising-elliptic,viswan2021,viswan-entropy,baxter}.

We begin by expressing the partition function of the Ising model on
selected planar lattices in terms of Mahler measures.
This approach yields hypergeometric formulas for the
partition functions.
Next, we establish a connection between the Ising model
at special temperatures and the Hasse-Weil $L$-function of an 
elliptic curve defined by the projective closure of the polynomial
that appears in the Mahler measure. 
Finally, at the critical point of the phase
transition,
we express 
the free
energy in terms of a Dirichlet
$L$-function.
The free energy is a physical quantity, whereas
$L$-functions are  number theoretic entities.  Why
should they
be  related?

{
By the projective closure of a curve in affine space, we mean its
  embedding in projective space by homogenizing the coordinates of its
  defining equation, which adds the missing points at infinity.
  Not every nonsingular cubic curve in affine space is elliptic, but
every nonsingular cubic curve becomes elliptic when considered in
projective space.
The term
{\it elliptic curve} conventionally means a nonsingular, projective, algebraic curve of genus one.
}

Section \ref{sec-mahler} introduces the basic facts regarding Mahler
measures, along with specific identities and hypergeometric formulas
that will be prerequisites for deriving the main results. Section
\ref{sec-ising} presents results for the Ising model on the square,
triangular and honeycomb lattices.  Concluding remarks are given in
Section~\ref{sec-concl}.

\section{Mahler measures.}
\label{sec-mahler}
The quantity that would later become known as the Mahler measure was
originally introduced in 1933 by Lehmer~\cite{lehmer}, who was
interested in finding large prime integers~\cite{zudilin-book}.  For a
monic polynomial, the Mahler measure is the product of the absolute
values of all the roots outside the complex circle.  More generally,
let \mbox{$p(x) = \sum_{i=0}^n a_i x^i = a_n \prod_{i=1}^n
  (x-\alpha_i)$.} Then the Mahler measure $M(p(x))$ is given by
\mbox{$M(p(x)) = |a_n| \prod_{i=1}^n \max(1,|\alpha_i|).$ }
The application of Jensen's formula leads to the equivalent (and
eponymous) analytic definition due to Mahler~\cite{mahler1962},
who  in the 1960s  used it 
to obtain  estimates in transcendence
 theory~\cite{zudilin-book,mahler-zeta},
$
 M(p(x)) =
\exp \left[
\frac 1 {2 \pi i }
\oint _{|z|=1} {\log |p(z)| }  {\D z \over z}
\right].
$
The logarithmic Mahler measure is defined as ${\mathrm m}(p(x))=\log
M(p(x))$. Following convention, ${\mathrm m}(p(x))$ is also known as
the Mahler measure, with the logarithm implicitly assumed. Mahler's
analytic definition generalizes to the multivariate case:
\be
{\mathrm m}(p(x_1,\ldots,x_d)) =
\frac 1 {(2 \pi i)^d }
\!\!\! \underset{{ {|z_1|=\ldots=|z_d|=1}}} {\!\!\!\!\oint\!\cdots\!\oint}
         {\!\!\!\!\!\!\log |p(z)| }  {\D z_1 \over z_1} \ldots{\D z_d \over z_d}
.
\nonumber
\ee

\begin{figure}[h]
  \vspace{0.5cm}
  \includegraphics[width=1\linewidth]{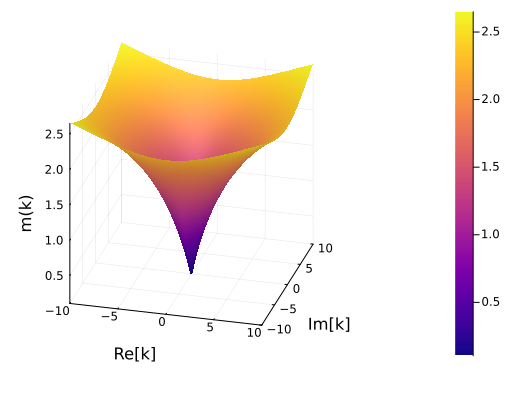}
\includegraphics[width=1\linewidth]{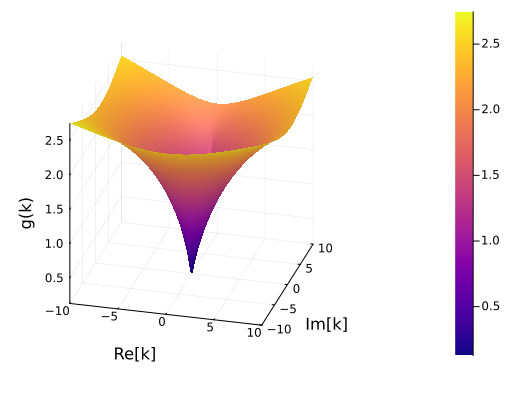}
\vspace{0.5cm}

\caption{Plots of $m(k)$ and $g(k)$ for $k\in \Bbb C$.
{  The differences between these plots
  can more clearly be seen in Figure~\ref{fig-m-g}.}
  The above 3D complex surface plots  were generated
    with code written in Julia for numerical integration using
    Gauss-Kronrod quadrature.}
  \label{fig-complex-m-g}.

\end{figure}

\begin{figure*}[t]
  \includegraphics[width=0.49\linewidth]{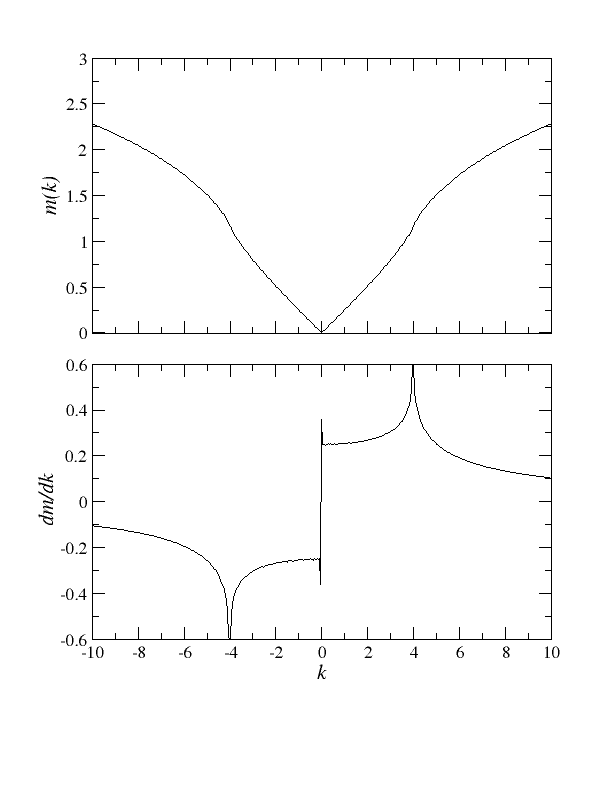}
  \includegraphics[width=0.49\linewidth]{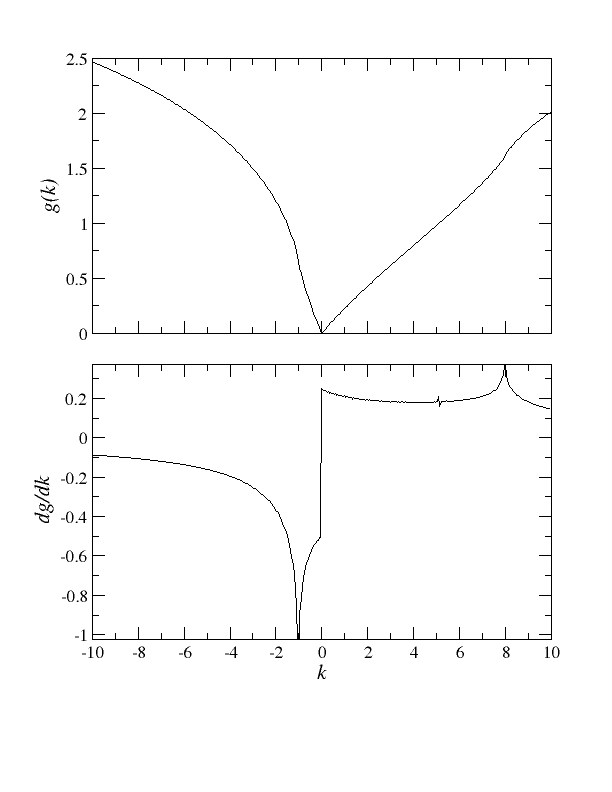}
\vspace{-10mm}
  \caption{Plots of $m(k)$ and $g(k)$ for real $k$ and their
    derivatives $dm/dk$ and $dg/dk$.  The singularities of $m(k)$ for
    $k\in\{0,\pm 4\}$ and of $g(k)$ for $k \in \{0,-1,8\}$ are clearly
    visible in the derivatives. For these singular values, the
    cubic curves defined by $P_1=0$ and $P_2=0$ are singular.
    As is well known in algebraic geometry, the
    projective closure of every nonsingular cubic curve over $\Bbb C$
    is an elliptic curve. }

\label{fig-m-g}
\end{figure*}

For our purposes related to the Ising model, we will need the
Mahler measures of two extensively researched Laurent polynomials in
the variables $x,y\in \mathbb  C$,
specifically~\cite{lalin2007,guttmann2012,g-singular},
\begin{align}
  P_1(x,y)&=k + x + 1/x + y + 1/y ~,
  \nonumber
  \\
  P_2(x,y)&= (x+ 1 ) (y+ 1)  ({x}  + y ) -  kxy~.
  \nonumber
\end{align}
The projective closures of the equations \mbox{$P_1(x,y)=0$} and
\mbox{$P_2(x,y)=0$} define elliptic curves for most values of the free
parameter $k \in \mathbb  C$~\cite{m-singular,g-singular}. The exceptions
are singular cubic curves when $k \in \{0,\pm 4\}$ for \mbox{$P_1=0$},
and similarly \mbox{$k\in\{-1,0,8\}$} for \mbox{$P_2=0$}.
The following Mahler measures are conventionally
known~\cite{guttmann2012} as $m$ and $g$:
\begin{align}
    m(k)&={\mathrm m}  (P_1) \label{eq-mm}~,\\
    g(k)&={\mathrm m}  (P_2) \label{eq-mg}~.
\end{align}
Plots are shown in Figs.~\ref{fig-complex-m-g} and \ref{fig-m-g}.
We will also need below, for the triangular and
honeycomb lattices, the following less well known Mahler measure: \be
\label{eq-t-k}
t(k) = {\mathrm m}(k + x + 1/x + y + 1/y + xy + 1/xy)~.
\ee

A remarkable relationship between Mahler measures and
\mbox{$L$-functions,} which inspired significant subsequent
research~\cite{boyd1981,boyd1998,smyth-book,zudilin-book,lalin2007,villegas-1999,deninger,rogers2010,g-singular,m-singular}, was discovered by  C.~J.~Smith
(and communicated by Boyd~\cite{boyd1981}) in 1981.
$L$-functions are of fundamental importance in number theory because
they link arithmetic properties with complex analysis.  The 
best known $L$-function is the Riemann zeta function
\mbox{$\zeta(s)= L(\chi_0^{(1)},s)$}, where $\chi^{(1)}_0$ is the principal
character modulo 1.
The
Dirichlet $L$-function $L(\chi,s)$
of a Dirichlet character $\chi(n)$ 
is  defined
according to
\be
L(\chi,s) = \sum_{n=1}^\infty  {\chi(n)\over n^s}~.
\nonumber
\ee
For the Ising model, 
we will need $L(\chi_{-3},s)$
  and  $L(\chi_{-4},s)$.
The real odd  Dirichlet character $\chi_{-f}$ of conductor $f$
can be written using  the Kronecker symbol as
$\chi_{-f} (n) = \left( { ~-f~ \over ~n~}
\right)$.
For example, when $f=3$ we have
\be
\nonumber
\chi_{-3}(n)
=
\left\{
\ba{ll}
0 & {\rm if~} n=0 \mod 3\\
1 & {\rm if~} n=1 \mod 3\\
-1 & {\rm if~} n=2 \mod 3~~~.\\
\ea
\right.
\ee 

A related $L$-function is the Hasse-Weil $L$-function $L(E,s)$ of an
elliptic curve $E$ over the rationals $\mathbb Q$, defined as follows.
Let $\chi(E,p)=1$ if the $E$ has good reduction at prime $p$ and
$\chi(E,p)=0$ otherwise.
{ 
  By bad reduction at prime $p$ is meant that the elliptic curve $E$
  over $\mathbb Q$, after reducing its coefficients modulo $p$ in
  the residue field $\mathbb F_p$, becomes singular in $\Bbb F_p$. If it
  remains nonsingular, then there is good reduction at prime $p$. A
  related concept is the conductor $N$ of an elliptic curve $E$.
  Specifically, $N$ is an integer that encodes the bad reduction of
  $E$ at various primes and is defined as the finite product of the
  prime powers $p^{N_p}$ for which the elliptic curve has bad
  reduction, where $N_p$ is determined by the type of bad reduction at
  each prime $p$.  For example, one can check that the curve $y^2 + x
  y + y = x^3 + x^2$ has bad reduction at 3 and 5, with the bad
  reduction being of the simplest kind, hence this curve has conductor
  $3\cdot 5$.

Let $a_p=p+1 - |E(\mathbb F_p)|$, where
$|E(\mathbb F_p)|$ is the number of points of $E$ modulo $p$. Then
the Hasse-Weil $L$-function is defined as}
\be
\nonumber
L(E,s)=   \prod_{{\mathrm {primes}~ }p} {L_p(E,s)}^{-1} =\sum_{n=1}^\infty {a_n \over n^s} ~,
\ee
where $ L_p(E,s)= 1- a_p p^{-s} + \chi(E,p) p^{1-2s}~.  $
In the 1990s, Deninger~\cite{deninger} interpreted
$\mathrm m(P)$
  for a Laurent polynomial $P$ over $\mathbb Q$ as a Deligne period of
  a mixed motive and conjectured that $m(1)$ could be expressed in
  terms of $L(E_{15},2)$. {A discussion of
    Deligne–Beilinson cohomology and Deninger's  method for
    multivariate polynomials is beyond the scope of this
    article. We refer the interested reader to
    ref.~\cite{zudilin-book}}.
Villegas~\cite{villegas-1999} studied $\mathrm m(P(k))$ as a function
of the parameter $k$ and its relation to modular forms and mirror
symmetry.
Boyd~\cite{boyd1998} argued that given a polynomial $P$ in two
variables, there should be relations such as
${\mathrm m}(P) = a L(E, 2)/\pi^2$, where $a \in \mathbb Q$.

There are many known formulas relating $m(k)$ and $g(k)$ to
$L$-functions of elliptic curves for special values of $k$. For
our purposes here related to the Ising model, it will suffice to
cite the following examples:

\begin{align}
  m(8)&={24 \over \pi^2} L(E_{24},2)~,
\label{eq-m8}
  \\
  g(-4)&= {18 \over \pi^2} L(E_{36},2) ~.
\label{eq-g-4}
\end{align}
Here, 
$E_{24}$ and $E_{36}$ are the elliptic curves of conductor $24$
and $36$ corresponding to $P_1=0$ and $P_2=0$, respectively.
These are Eqs.~(27) and (17)  listed
by Rogers~\cite{rogers2010} (or (2.19) and {(2.9)} in the arXiv version).
Many such formulas have been proven~\cite{lalin2007,g-singular,m-singular,rogers2010}.

Relationships represented by equations such as (\ref{eq-m8}) and
(\ref{eq-g-4}) are considered deep because the
Mahler measure on the left-hand side is a function only of the
polynomial --- not of the elliptic curve that it
defines~\cite{boyd1998}. In contrast, the $L$-function $L(E,s)$
depends on the elliptic curve itself.  See
refs.~\cite{smyth-book,zudilin-book,boyd1998} for details.

We will see below that the critical temperatures of the Ising model on our 
chosen lattices are related to the singularities in $m(k)$ and $g(k)$
at \mbox{$k=4$} and \mbox{$k=8$} respectively.
It is known~\cite{guttmann2012} that $m(4)$ and $g(8)$ can be
expressed in terms of Dirichlet $L$-functions,
\begin{align}
  m(4)&= \label{eq-m4}
{4  \over \pi} L(\chi_{-4}, 2)= {4
  G \over \pi}~,\\
g(8) &=
{15 \sqrt 3 \over 4\pi}
L(\chi_{-3}, 2) ~, \label{eq-g8}
\end{align}
where $G$ is the Catalan constant.

Formulas such as (\ref{eq-m8})--(\ref{eq-g8}) are known only for
special values of $k\in \mathbb  Q$, whereas hypergeometric formulas are
known for arbitrary $k\in \mathbb  C$.  Let $_pF_q$ denote, as usual, the
generalized hypergeometric function:
\be
\nonumber
_pF_q\left[
  \ba{c}{a_1,a_2,\dots, a_p} \\ {b_1,b_2,\dots,b_q}\ea ;x \right] =
\sum_{n=0}^\infty {(a_1)_n(a_2)_n \dots (a_p)_n \over (b_1)_n(b_2)_n
  \dots (b_q)_n } ~{x^n \over {n!}}.
\ee
The Pochhammer symbol $ (x)_n = {\Gamma(x+n) / \Gamma(x)}$ is
the rising factorial. One can express
$m(k)$ and, for $|k|$ sufficiently large, 
$g(k)$ as,
\begin{align}
m(k) &= \Re  \left(\log k  - {2\over k^2}   
~{_4}F_3
\left[ \ba{c}{1,1,{3\over 2},{3\over 2}} \\ {2,2,2}\ea 
; {16 \over k^2 } \right]  \right) ~,
\label{eq-m-h}\\
  g(k) &  = \frac 1 3 \nonumber
\Re
\bigg\{
  \log\left( (4+k) (k-2)^4 \over k^2 \right)
\\& \quad\quad \nonumber
  - {2k^2 \over (4+k)^3}~ _4F_3
\left[ \ba{c}{1,1,{4\over 3},{5\over 3}} \\ {2,2,2}\ea 
  ; {27k^2 \over (4+k)^3}  \right]
\\
& \quad\quad
  - {8k \over (k-2)^3}~ _4F_3
\left[ \ba{c}{1,1,{4\over 3},{5\over 3}} \\ {2,2,2}\ea 
  ; {27k \over (k-2)^3}  \right]  \bigg\}~.
\label{eq-g-h}
\end{align}
Eq.~(\ref{eq-m-h}) for $m(k)$ is well known
~\cite{zudilin-book,guttmann2012,villegas-1999,lalin2007}. It can be
obtained by using \mbox{$\log |z|= \Re (\log z)$,} where $\Re(z)$ is
the real part of $z$, expanding the logarithm in the Mercator series,
applying the binomial theorem, and then integrating term by term.
The formula for $g(k)$ can be approached using
$q$-series~\cite{lalin2007}.  Specifically, Eq. (\ref{eq-g-h}) can be proven using a
modular expansion \cite{stienstra-dimer}, see for instance Eq.(2-37)
of \cite{lalin2007} or Eq.~(17) of \cite{guttmann2012}.
We did not find in the literature~\cite{glasser-of-glasser} a
hypergeometric formula for $t(k)$, which we will need below to treat
the honeycomb and triangular lattices.  We attempted to calculate
$t(k)$ using the Mercator series, however the resulting expression
involves multiple sums that resisted simplification.  (See
Supplementary  Material S1
for the full derivation.)
In contrast to the straightforward calculation of $m(k)$ (detailed in
Supplementary Material S2 for comparison), the brute force method
described above is unable to yield the hypergeometric formula for
$t(k)$.  As a result, we reach an impasse.

A different, more creative, approach is needed for $t(k)$.  The method
below was inspired by ref.~\cite{guttmann2012}.  It is easy to check
explicitly that at each step below any change of variables leaves the
Mahler measure invariant:
\begin{align}
  t(k)
 &= \nonumber 
  \mathrm m ( k + xy + 1/(xy) + x/y+y/x + x^2+ 1/x^2 )
  \\
  &=
  \mathrm m ( k-2 + (x+1/x)(y+1/y + x+1/x)) \nonumber 
\\
& 
= {\mathrm m} \left(
k -2 + \left(x^2+ 1/{x^2}\right) \left(xy + 1/{xy}\right)  \left( x/ y + y/ x\right)
\right)
\nonumber
\\
&= \nonumber
 {\mathrm m} \left(
k -2 + \left(x^2+ 1/{x^2}\right) \left(y+1/{y}\right)  \left( {x^2}/ y +  y/ {x^2}\right)
\right)
\\
&=
{\mathrm m} \left(
  (x^4+ 1 ) (y^2+ 1)  ({x^4}  + y^2 ) - (2 -k)x^4y^2
\right) \nonumber
\\
&=
{\mathrm m} \left(
  (x+ 1 ) (y+ 1)  ({x}  + y ) - (2 -k)x y
\right) . %
\end{align}
          {Finally, compare with (\ref{eq-mg}) and $P_2$ 
to
obtain
\be
\label{eq-t-g}
t(k)=  g(2-k)~. %
\ee
            This result is not new and has been known in slightly different
form~\cite{boyd1998,lalin2013}.
See Supplementary Material S3 for further details.}

We will use 
the above relation between Mahler measures  
to reformulate the Ising model on the triangular
and honeycomb lattices.
The  Mahler measure corresponding to the partition function of the
Ising model on the square lattice was investigated in the 2010s
~\cite{guttmann2012,viswan2017-pre}, leading to hypergeometric
formulas~\cite{hucht2011,viswan2021,viswan2017-pre}.  However, even
for the square lattice, no connection to $L$-functions via Mahler
measures seems to have been previously recognized. Notably, Mahler
measures were not used in some of the first studies of the links
between the Ising model, elliptic curves and $L$-functions.  In 2001,
correlations of the Ising model on the square lattice were studied via
$L$-functions of finite graphs~\cite{ising-l-function}. A few years
later it shown that
linear differential operators associated with the susceptibility
of the Ising model are associated with
elliptic curves, with an interpretation of complex multiplication for
elliptic curves as (complex) fixed points of generators of the exact
renormalization group~\cite{ising-sigma,ising-elliptic}.  
Below we present results that significantly advance beyond the known
results.

\begin{figure*}[t]
  \includegraphics[width=0.45\linewidth]{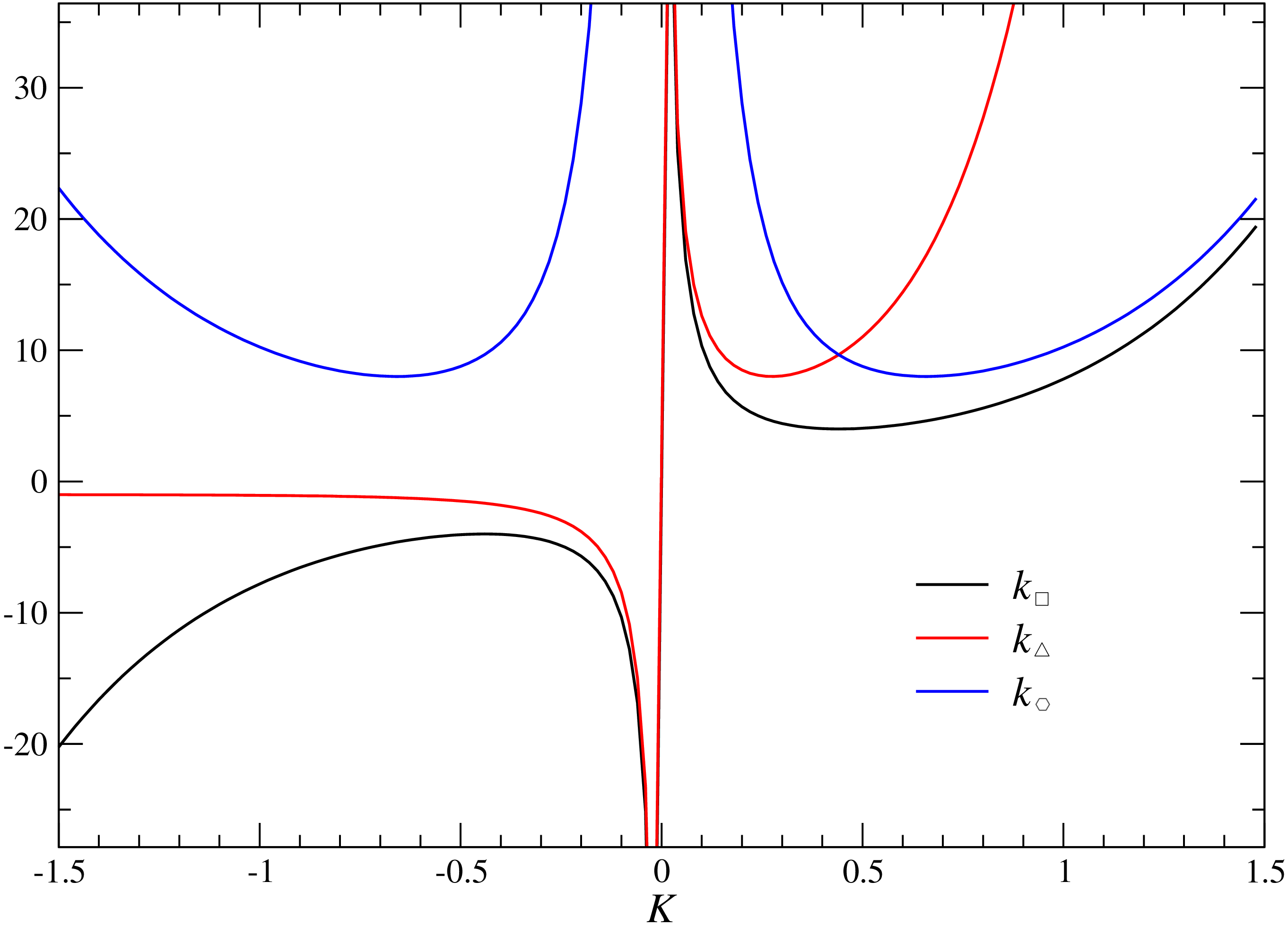}~~~~~~
  \includegraphics[width=0.45\linewidth]{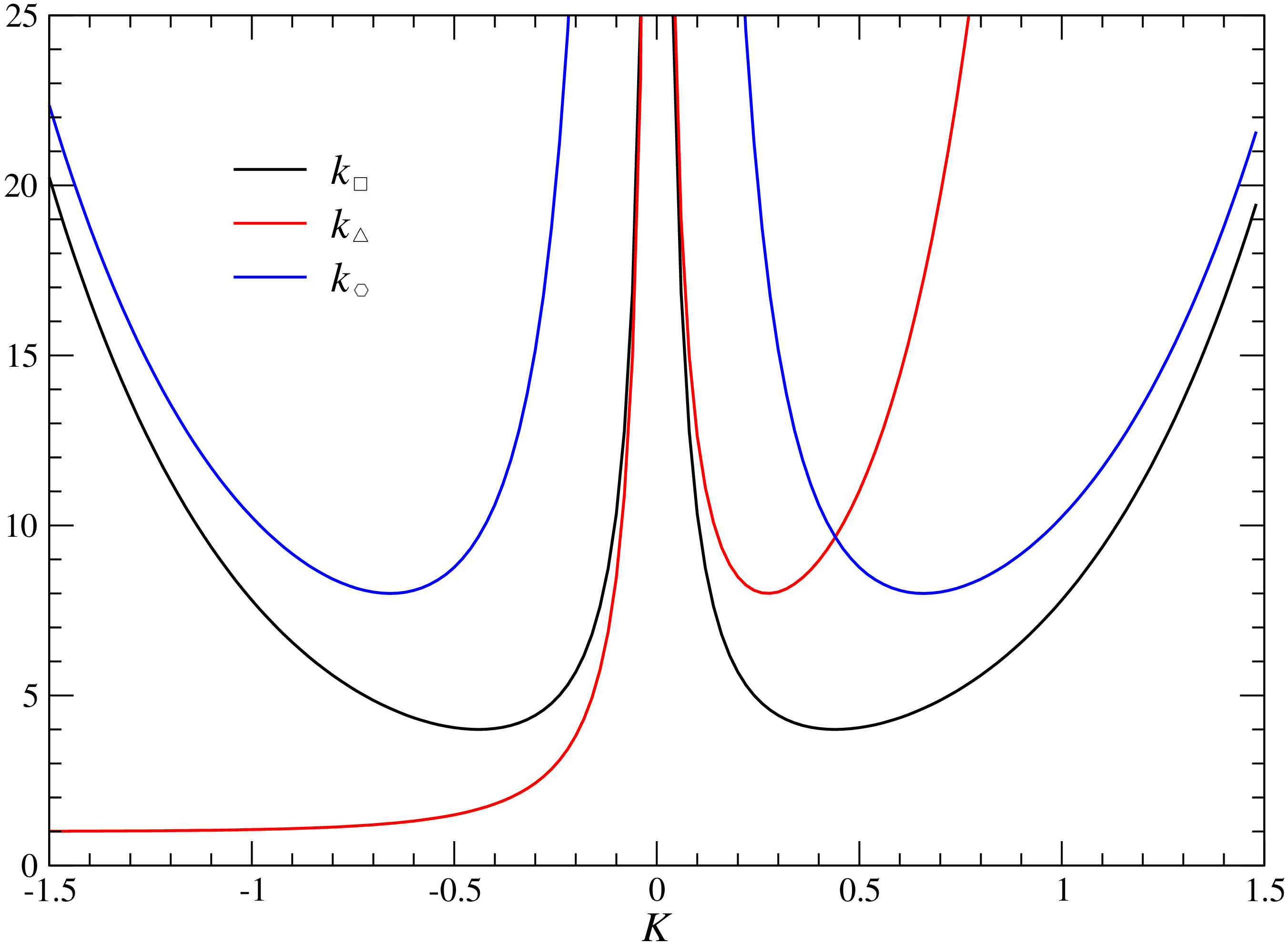}

\caption{Plots of $k_\square$, $k_\triangle$ and $k_{\small\hexagon}$
  (left) and their absolute values (right) as functions of the reduced
  temperature $K$. Note that $k_\triangle \to -1$ as $K\to
  -\infty$. The hypergeometric formula for $g(k)$ does not converge to
  $g(k)$ for small $|k_\triangle|$, which affects Eq.~(\ref{eq-tri-h}).
  This behavior on the triangular
  lattice is due to frustration, i.e.  the fact that for negative $K$,
  it is not possible for all 3 nearest neighbor spins on a triangle be
  pairwise anti-aligned. In contrast, on both the square and honeycomb
  lattices there is the symmetry $Z(-K)=Z(K)$.}

\label{fig-k}
\end{figure*}

\section{The Ising model}
\label{sec-ising}
 We next use Mahler measures to establish a direct (and unforeseen)
link between the free energy of the Ising model, elliptic curves and
$L$-functions.
To this end, we begin by defining 
{ the Hamiltonian of the ferromagnetic Ising model with $N$ classical
  spins with isotropic nearest-neighbor interactions and zero external
  magnetic field as $ H_N= - J \sum_{\exval{ij}}
  \sigma_i \sigma_j $.
Here  $J$ is the isotropic coupling, $\sigma_i=\pm 1$ and
$\exval{ij}$  the set of all pairs $(i,j)$ of nearest-neighbor
``spins'' on the chosen lattice.  
We restrict our attention below  to the square, triangular
and honeycomb lattices, denoted where necessary by the symbols
$\square,~\triangle$ and {\large \hexagon} respectively.
Let
\mbox{$\beta=1/(k_{\mbox{\tiny B}}T)$} denote 
the temperature parameter and
\mbox{$K= \beta J$} the reduced temperature.  The canonical
partition function is given by \mbox{$Z_N(K)= \sum \exp[-\beta H_N]$,}
where the sum is over all possible states. The per-site free energy
$f(K)$ is given by {$-\beta f(K)= \log Z(K)$}, where $Z$ is the per-site
partition function {$Z(K) = \lim_{N\to \infty}[Z_N(K)]^{1/N}$.}  }

The partition functions in the infinite square, triangular and
hexagonal lattices are given by
{
\begin{align}
\log {Z_{\square}(K)\over  2} &=   \nonumber
\frac{1}{2 \pi^2}  
\iint_0^{\pi}   
\log[\cosh^2 2K \\ &  -\sinh 2K ( \cos x + \cos y )]
\label{eq-ons}
~\D x \D y~,\\
  \log{ Z_{\triangle}(K)\over 2} &
= \frac 1 {8\pi^2} \nonumber
\iint_0^{2\pi}  
\log [\cosh^3 2K + \sinh^3 2K
\\ & 
  -\sinh 2K~(\cos x + \cos y + \cos(x+y))] ~\D x\D y ~,\\
\log {Z_{\hexagon}(K) \over 2} &
=
\frac 1 {16\pi^2}
\iint_0^{2\pi}   \nonumber 
\log \tfrac 1 2 [\cosh^3 2K +1 \\ &  - \sinh^2 2K~
(\cos x + \cos y + \cos(x+y))] \,\D x\D y \,.
\end{align}
}
Here (\ref{eq-ons}) is Onsager's solution~\cite{onsager}, while the
triangular and honeycomb lattices were solved in \cite{hexagon}.  The
critical temperatures are $K_c^\square = \tfrac{1}{2} \log
\left(\sqrt{2}+1\right)$, $K_c^\triangle = \tfrac{1}{4} \log (3)$
and $K^{\hexagon}_c = 
\tfrac{1}{2} \log \left(\sqrt{3}+2\right)$.

We next rewrite these partition function in terms of Mahler measures.
For the square lattice, we will use $m(k)$. For the triangular and
honeycomb lattices, we will use $t(k)$ and later apply
(\ref{eq-t-g}).
Noting \mbox{$Z(0)=2$} for all three lattices, 
for nonzero real $K \in \mathbb  R \bs \{0\}$, we get
{
\begin{align}
&\! \log {Z_{\square}(K)\over 2} =  \! \nonumber \frac 1 2\left(  \log | \tfrac 1 2 \sinh 2K| 
+
m(2\cosh 2K \coth 2K)\right)~,
  \\[5mm]
&  \!    \log {Z_{\triangle}(K) \over 2} = \nonumber
  \frac 1 2  \bigg(   \log |\tfrac 1 2 \sinh 2K|
  \\ \nonumber
  &
  \quad  \quad  \quad  \quad  \quad   \quad  \quad  \quad
  + t\left(-
    {2 \cosh^3  2K + 2\sinh^3 \!2K \over \sinh 2K} 
\right)     \bigg)~,
  \\[5mm]
&   \!   \log  {Z_{\hexagon}(K) \over 2^{1/2}} =\!
  \frac 1 4  \!\left( \!  2 \log |\sinh 2K| \!+\! t\left(\!-
    {2\cosh^3 2K + 2  \over \sinh^2 2K}
    \right) \!    \right)~.
  \label{eq-ereirhgiuerhgiuerhg}
\end{align}}
  Let 
\begin{align}
k_{\square} &=   2\cosh 2K \coth 2K~, \\
k_{\triangle} &=  2+
{2\cosh^3 2K + 2\sinh^3 2K \over \sinh 2K} \nonumber \\
&=  2 \cosh ^2(2 K) (\coth (2 K)+1)~,
\\
  k_{\hexagon} &=
  2+ {2\cosh^3 2K + 2  \over \sinh^22K} =
\cosh ^2(2 K) \text{csch}^2(K) ~.
\end{align}
Then from (\ref{eq-ereirhgiuerhgiuerhg})
and applying (\ref{eq-t-g})
we get{
\begin{align}
& \log {Z_{\square}(K)\over 2} =   \nonumber \frac 1 2\left(  \log |\tfrac 1 2 \sinh 2K| 
+
m( k_{\square}) \right)~,
  \\
&      \log {Z_{\triangle}(K) \over 2} = \nonumber
  \frac 1 2  \left(   \log |\tfrac 1 2 \sinh 2K| + g(k_{\triangle})
\right)~,
  \\
&      \log  {Z_{\hexagon} (K)\over 2^{1/2}} =
  \frac 1 4  \left(   2 \log |\sinh 2K| + g(  k_{\hexagon})    \right)~.
  \label{eq-effreirhgiuerhgiuevvrhg}
\end{align}
}
See Fig.~\ref{fig-k}.

Hypergeometric formulas follow from (\ref{eq-m-h}) and (\ref{eq-g-h}):
{
\begin{align}
&\log  Z_{\square} (K) 
= \log(2 \cosh 2K) 
- {1\over k_{\square}^2}
~{_4}F_3
\left[ \ba{c}{1,1,{3\over 2},{3\over 2}} \\ {2,2,2}\ea 
; {16 \over k_{\square}^2} \right]  ,
\label{eq-sq-h}
\\
&  \log {Z_{\triangle}(K) \over 2^{1/2}} = \nonumber
\frac 1 2   \log| \sinh 2K|
\\ &
\quad
\quad
+
 \frac 1 6 \nonumber
  \log\left( (4+k_{\triangle}) (k_{\triangle}-2)^4 \over k_{\triangle}^2 \right)
\\& \quad \quad
\nonumber
  - {2k_{\triangle}^2 \over 2(4+k_{\triangle})^3}~ _4F_3
\left[ \ba{c}{1,1,{4\over 3},{5\over 3}} \\ {2,2,2}\ea 
  ; {27k_{\triangle}^2 \over (4+k_{\triangle})^3}  \right]
\\
& \quad\quad
  - {8k_{\triangle} \over 2(k_{\triangle}-2)^3}~ _4F_3
\left[ \ba{c}{1,1,{4\over 3},{5\over 3}} \\ {2,2,2}\ea 
  ; {27k_{\triangle} \over (k_{\triangle}-2)^3}  \right] ~,
\label{eq-tri-h}
\\[5mm]
& \log{Z_{\hexagon}(K) \over 2^{1/2}} = \nonumber
{\frac 1 2 \log |\sinh 2 K|}
\\
& \quad\quad
+
 \frac 1 {12} \nonumber
  \log\left( (4+k_{\hexagon}) (k_{\hexagon}-2)^4 \over k_{\hexagon}^2 \right)
\\& \quad\quad
\nonumber
  - {k_{\hexagon}^2 \over 2(4+k_{\hexagon})^3}~ _4F_3
\left[ \ba{c}{1,1,{4\over 3},{5\over 3}} \\ {2,2,2}\ea 
  ; {27k_{\hexagon}^2 \over (4+k_{\hexagon})^3}  \right]
\\
& \quad\quad
  - {2k_{\hexagon} \over (k_{\hexagon}-2)^3}~ _4F_3
\left[ \ba{c}{1,1,{4\over 3},{5\over 3}} \\ {2,2,2}\ea 
  ; {27k_{\hexagon} \over (k_{\hexagon}-2)^3}  \right] ~.
\label{eq-hex-h}
\end{align}
}

\begin{figure}[t]
  \includegraphics[width=1\linewidth]{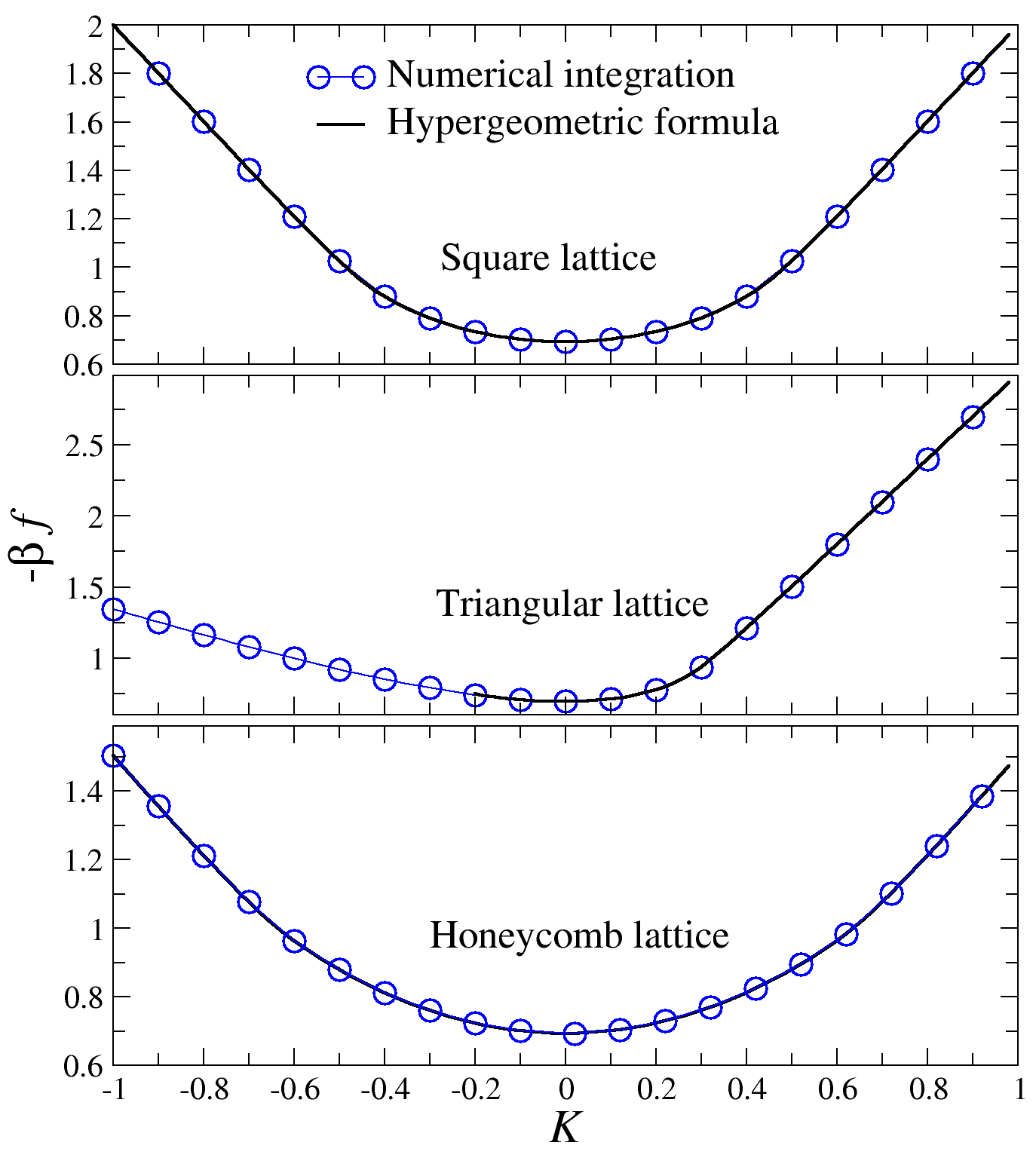}
  \vspace{-4mm}
  \caption{Comparison of $-\beta f = \log Z$ for the square, triangular
  and honeycomb lattices calculated via numerical integration and via
  the hypergeometric formulas.  }
\label{fig-z}
\end{figure}

See Figure~\ref{fig-z}.    At sufficiently negative temperatures,
  the hypergeometric expression for $Z_{\triangle}$ fails because
  $|k_{\triangle}|$ becomes too small (see Fig.~\ref{fig-k}).
Series (\ref{eq-sq-h}) is not new, as discussed below.

Next, we obtain a connection with $L$-functions of elliptic
curves, using (\ref{eq-m8}) and
(\ref{eq-g-4}) as illustrative examples. 
Let
$K'_{\square},~K'_{\triangle}$ and
$K'_{\hexagon}$ be the temperatures corresponding to
\mbox{$k_{\square}=8$} and
\mbox{$k_{\triangle}=k_{\hexagon}=-4$.}
The choice \mbox{$k_{\hexagon} =-4$} does not correspond to
temperatures \mbox{$K_{\hexagon}\in \mathbb  R$.}  Since our goal is to
show the connection with $L$-functions, we can obtain the required
complex valued temperature \mbox{$K_{\hexagon}\in \mathbb  C$} via
analytic continuation. Only one of several complex-valued temperatures
is shown.  For the square lattice, there are 2 real-valued
temperatures corresponding to \mbox{$k_{\square}=8$}, only one of
which we show here for illustration purposes.  For the triangular
lattice, the reduced temperature $K=\beta J$ is negative, which can be
achieved for $\beta>0$ by choosing $J<0$. The values thus obtained are
\begin{align}\nonumber
K'_{\square} &= \frac{1}{2} \log \left(\sqrt{6}-\sqrt{2
  \sqrt{6}+5}+2\right) \approx 0.132~,\\
\nonumber
K'_{\triangle} & =
\frac{1}{4} \log \left(2 \sqrt{3}-3\right)\approx -0.19~,
\\
\nonumber
K'_{\hexagon} &=
\log \left(-\sqrt{\sqrt{3}-i \sqrt{2 \sqrt{3}-3}-1}\right)
\approx 2.77 i~.
\end{align}
Substituting these and (\ref{eq-m8}) and (\ref{eq-g-4}),
into (\ref{eq-effreirhgiuerhgiuevvrhg}), we obtain a direct 
connection between the free energy and $L$-functions of elliptic
curves:
\begin{align}
  -\beta f_{\square}(K'_{\square}) &=
\frac 1 2\left(  \log |\tfrac 1 2 \sinh 2K' _{\square}|
+ {24\over \pi^2} L(E_{24},2)\right)\nonumber \\ & \quad\quad + \log 2
~,   \nonumber
  \\
-\beta f_\triangle (K'_\triangle) 
&=
\frac 1 2  \left(   \log| \tfrac 1 2 \sinh 2K'_\triangle |
+ {18\over \pi^2} L(E_{36},2)
\right) \nonumber \\ & \quad\quad + \log 2~,   \nonumber
\\
  -\beta f_{\hexagon} (K'_{\hexagon})  
& =
            {  \frac 1 4  \left    (  2 \log| \sinh 2 K'_{\hexagon}|   + {18\over \pi^2} L(E_{36},2) \right)}
  \nonumber
  \\ & \quad\quad + \tfrac 1 2 \log 2~.
\label{eq-elliptic-sth}
\end{align}

We finally present
our last result,
unveiling a connection that might be even deeper.
The positive critical
temperatures are given by
the singularities at 
\mbox{$k_\triangle=k_{\hexagon}=8$} and
\mbox{$k_{\square}=4$}.  Using (\ref{eq-m4}) and (\ref{eq-g8}) in
(\ref{eq-effreirhgiuerhgiuevvrhg}) and after simplification, we arrive
at the critical free energy as Dirichlet
$L$-functions $L(\chi,s)$ at $s=2$:
\begin{align}
    \nonumber
-\beta f_{\square}(K_c^\square) &=   \frac 1 2 \log 2 ~ +~  {2  \over\pi}~ L(\chi_{-4}, 2)
~, %
\\[8pt]
  \nonumber
-\beta f_\triangle (K_c^\triangle) 
&=   \frac 1 4 \log {4\over 3} +  {15 \sqrt 3 \over 8\pi}
L(\chi_{-3}, 2) ~, %
\\[8pt]
-\beta f_{\hexagon} (K_c^{\hexagon})  
& =   \frac 1  4\log {12       }
+  {15 \sqrt 3 \over 16\pi}
L(\chi_{-3}, 2)
~. 
\label{eq-f-sth}
\end{align}
Onsager's critical  free energy for the square lattice (see Eq. (121) in
~\cite{onsager}) is usually written in terms of $G$, whereas in
(\ref{eq-f-sth}) we have highlighted instead the $L$-function.

\section{Concluding remarks.}
\label{sec-concl}
We have shown that Mahler measures can serve as a bridge connecting
the Ising model with elliptic curves and $L$-functions.
Eqs.~(\ref{eq-elliptic-sth}) and (\ref{eq-f-sth})
unveil
unexpected relationships for the free energy.
We emphasize that the latter is a physical quantity, while
$L$-functions originate in number theory.  They are fundamentally
different --- yet intriguingly related.  As Poincaré once
remarked~\cite{poincare}, the most valuable relationships are those
that
``reveal unsuspected relations between other facts, long since known,
but wrongly believed to be unrelated to each other.''
We hope that the findings presented above   motivate others to further
explore these and similar relationships.

Other advances include the reformulation
(\ref{eq-effreirhgiuerhgiuevvrhg}) in terms of Mahler measures, and
similarly the Hypergeometric formulas (\ref{eq-tri-h}) and
(\ref{eq-hex-h}) for the triangular and honeycomb lattices. In
contrast, (\ref{eq-sq-h}) has been known for more than a
decade~\cite{hucht2011,viswan2021,viswan2017-pre}. It is noteworthy
that Glasser and Onsager had derived a (different) hypergeometric
formula in the 1970s that they never published~\cite{viswan2021}.

Finally, we discuss the main limitations of this work and related open
problems:
(i) We have given only one example of a Hasse-Weil $L$-function for
each of the three lattices, while the list in 
ref.~\cite{rogers2010} yields  several others.
(ii) Only three lattices have been studied.  Can we generalize to
other planar lattices?  The $Z$-invariant Ising
model~\cite{baxter,yang2002} might be a promising candidate for
further study.
(iii) What of nonplanar lattices?  Intriguingly, while spanning tree
constants on planar lattices are related to $L$-functions $L(\cdot,s)$
evaluated at \mbox{$s=2$}, it is known~\cite{guttmann2012} that on the
simple cubic lattice the spanning tree constant can be expressed in
terms of an $L$-function evaluated at \mbox{$s=3$.} Might the free
energy of the critical Ising model on the simple cubic lattice similarly be
given by an $L$-function at \mbox{$s=3$~?}~
Such fascinating questions are currently unanswerable because the 3D
Ising model remains
unsolved~\cite{viswan-entropy,perk1,perk2,perk3,perk4}.
Exploring fresh ideas will be essential for
advancing our understanding of these formidable challenges.

\bigskip
\section*{Acknowledgements}

We thank J. H. H. Perk for discussions via e-mail over a period of
several years. { The anonymous referees gave very helpful
  comments. } This work was supported by CNPq (Grant No.
302414/2022-3).

\clearpage



\section*{Supplementary material}

# Supplementary Material for:

# Mahler measures, elliptic curves and $L$-functions for the free energy of the Ising model

Gandhimohan M. Viswanathan

*Department of Physics and National Institute of Science and Technology of Complex Systems,
Universidade Federal do Rio Grande do Norte, 59078-970 Natal–RN, Brazil*

### S1. DERIVATION OF A SUM FORMULA FOR $t(k)$

In this exercise we derive a sum formula for $t(k)$. We proceed starting with (3), with $\delta_i^j$ denoting the Kronecker delta function.

$$t(k) = \frac{1}{4\pi^2} \iint_0^{2\pi} \log|k + 2\cos x + 2\cos y + 2\cos(x+y)| \, dx dy \tag{S1}$$

$$= \Re\left[\frac{1}{4\pi^2} \iint_0^{2\pi} \log(k + 2\cos x + 2\cos y + 2\cos(x+y)) \, dx dy\right] \tag{S2}$$

$$= \Re\left[\log k + \frac{1}{4\pi^2} \iint_0^{2\pi} \log(1 + 2(\cos x + \cos y + \cos(x+y))/k) \, dx dy\right] \tag{S3}$$

$$= \Re\left[\log k + \frac{1}{4\pi^2} \iint_0^{2\pi} \log\left(1 + \frac{e^{ix} + e^{-ix} + e^{iy} + e^{-iy} + e^{i(x+y)} + e^{-i(x+y)}}{k}\right) dx dy\right] \tag{S4}$$

$$= \Re\left[\log k - \frac{1}{4\pi^2} \iint_0^{2\pi} \sum_{n=1}^{\infty} \frac{(-1)^n}{n} \left(\frac{e^{ix} + e^{-ix} + e^{iy} + e^{-iy} + e^{i(x+y)} + e^{-i(x+y)}}{k}\right)^n dx dy\right] \tag{S5}$$

$$= \Re\left[\log k - \frac{1}{4\pi^2} \iint_0^{2\pi} \sum_{n=1}^{\infty} \frac{(-1)^n}{nk^n} \sum_{j_i, i=1\ldots 6} \binom{n}{j_1 \ j_2 \ j_3 \ j_4 \ j_5 \ j_6} e^{i(j_1 x - j_2 x + j_3 y - j_4 y + j_5(x+y) - j_6(x+y))} dx dy\right] \tag{S6}$$

$$= \Re\left[\log k - \frac{1}{4\pi^2} \sum_{n=1}^{\infty} \frac{(-1)^n}{nk^n} \sum_{j_i, i=1\ldots 6} \binom{n}{j_1 \ j_2 \ j_3 \ j_4 \ j_5 \ j_6} \iint_0^{2\pi} e^{i(x(j_1 - j_2 + j_5 - j_6) + y(j_3 - j_4 + j_5 - j_6))} dx dy\right] \tag{S7}$$

$$= \Re\left[\log k - \sum_{n=1}^{\infty} \frac{(-1)^n}{nk^n} \sum_{j_i, i=1\ldots 6} \binom{n}{j_1 \ j_2 \ j_3 \ j_4 \ j_5 \ j_6} \delta^{j_2}_{j_1+j_5-j_6} \delta^{j_4}_{j_3+j_5-j_6}\right] \tag{S8}$$

$$= \Re\left[\log k - \sum_{n=1}^{\infty} \frac{(-1)^n}{nk^n} \sum_{j_1, j_3, j_5, j_6} \binom{n}{j_1 \ j_2 \ j_3 \ j_4 \ j_5 \ j_6}\right] . \tag{S9}$$

Next, we use the fact that

$$n = \sum_{i=1}^{6} j_i = j_1 + (j_1 + j_5 - j_6) + j_3 + (j_3 + j_5 - j_6) + j_5 + j_6 \tag{S10}$$

$$= 2j_1 + 2j_3 + 3j_5 - j_6 , \tag{S11}$$

so that

$$j_6 = 2j_1 + 2j_3 + 3j_5 - n , \tag{S12}$$

$$j_2 = j_1 + j_5 - j_6 = -j_1 - 2j_3 - 2j_5 + n , \tag{S13}$$

$$j_4 = j_3 + j_5 - j_6 = -2j_1 - j_3 - 2j_5 + n . \tag{S14}$$

We thus get

$$t(k) = \Re\left[\log k - \sum_{n=1}^{\infty} \frac{(-1)^n}{nk^n} \sum_{j_1,j_3,j_5} \binom{n}{j_1 \ (n-j_1-2j_3-2j_5) \ j_3 \ (n-2j_1-j_3-2j_5) \ j_5 \ (2j_1+2j_3+3j_5-n)}\right]. \tag{S15}$$

Changing variables $(j_1, j_3, j_5) \to (j_1, j_2, j_3)$ and switching to factorials, we get

$$t(k) = \Re\left[\log k - \sum_{n=1}^{\infty} \frac{(-1)^n}{nk^n} \sum_{j_1,j_2,j_3} \frac{n!}{j_1! \ (n-j_1-2j_2-2j_3)! \ j_2! \ (n-2j_1-j_2-2j_3)! \ j_3! \ (2j_1+2j_2+3j_3-n)!}\right] \tag{S16}$$

$$= \Re\left[\log k - \sum_{n=1}^{\infty} \frac{(-1)^n(n-1)!}{k^n} \sum_{j_1,j_2,j_3=0}^{n} \frac{1}{j_1!(n-j_1-2j_2-2j_3)!j_2!(n-2j_1-j_2-2j_3)!j_3!(2j_1+2j_2+3j_3-n)!}\right]. \tag{S17}$$

The sums in $j_{1,2,3}$ can run from $0 \ldots n$ without worry about satisfying (S10) because any factorial of a negative integer in the denominator is infinite.

As explained in the main text, we have not been able to simplify the sum, thus motivating the alternative approach via (11).

## S2. DERIVATION OF THE HYPERGEOMETRIC FORMULA FOR $m(k)$

Eq. (8) can be found in many references. What follows in this section is merely for comparison with formula (S17) for $t(k)$. We present a simple step-by-step derivation merely as an exercise and a convenient reference for readers who may wish to see all the details. We will first rewrite $m(t)$ in the form

$$m(t) = \frac{1}{(2\pi)^2} \iint_0^{2\pi} \log|k + e^{ix} + e^{-ix} + e^{iy} + e^{-iy}| dxdy \ . \tag{S18}$$

$$= \frac{1}{(2\pi)^2} \Re \iint_0^{2\pi} \log(k + e^{ix} + e^{-ix} + e^{iy} + e^{-iy}) dxdy \ . \tag{S19}$$

$$= \Re\left[\log k + \frac{1}{(2\pi)^2} \iint_0^{2\pi} \log(1 + (e^{ix} + e^{-ix} + e^{iy} + e^{-iy})/k) dxdy\right] . \tag{S20}$$

Next we expand the logarithm in the Mercator series, followed by the multinomial theorem:

$$m(t) = \Re\left[\log k - \frac{1}{(2\pi)^2} \iint_0^{2\pi} \sum_{n=1}^{\infty} \frac{(-1)^n}{k^n n} (e^{ix} + e^{-ix} + e^{iy} + e^{-iy})^n dxdy \ .\right] \tag{S21}$$

$$= \Re\left[\log k - \sum_{n=1}^{\infty} \frac{(-1)^n}{k^n n} \frac{1}{(2\pi)^2} \iint_0^{2\pi} (e^{ix} + e^{-ix} + e^{iy} + e^{-iy})^n dxdy\right] \tag{S22}$$

$$= \Re\left[\log k - \sum_{n=1}^{\infty} \frac{(-1)^n}{k^n n} \frac{1}{(2\pi)^2} \iint_0^{2\pi} \sum_{j_1,j_2,j_3,j_4=0} \binom{n}{j_1 \ j_2 \ j_3 \ j_4} e^{i(j_1 x - j_2 x + j_3 y - j_4 y)} dxdy\right] \tag{S23}$$

$$= \Re\left[\log k - \sum_{n=1}^{\infty} \frac{(-1)^n}{k^n n} \sum_{j_1,j_2,j_3,j_4=0} \binom{n}{j_1 \ j_2 \ j_3 \ j_4} \frac{1}{(2\pi)^2} \iint_0^{2\pi} e^{i(j_1 x - j_2 x + j_3 y - j_4 y)} dxdy\right] \tag{S24}$$

$$= \Re\left[\log k - \sum_{n=1}^{\infty} \frac{(-1)^n}{k^n n} \sum_{j_1,j_3=0} \binom{n}{j_1 \ j_1 \ j_3 \ j_3}\right] \tag{S25}$$

$$= \Re\left[\log k - \sum_{n=1}^{\infty} \frac{(-1)^n}{k^n n} \sum_{j_1,j_3=0} \frac{n!}{(j_1!)^2 (j_3!)^2}\right] \tag{S26}$$

Odd powers in $n$ vanish because we require $2j_1 + 2j_3 = n$. So we get

$$m(t) = \Re\left[\log k - \sum_{n=1}^{\infty} \frac{1}{k^{2n} 2n} \sum_{j=0} \frac{(2n)!}{(j!)^2((n-j)!)^2}\right] \tag{S27}$$

$$= \Re\left[\log k - \sum_{n=1}^{\infty} \frac{1(2n-1)!}{k^{2n}} \sum_{j=0} \frac{1}{(j!)^2((n-j)!)^2}\right] \tag{S28}$$

$$= \Re\left[\log k - \sum_{n=1}^{\infty} \frac{1(2n-1)!}{k^{2n}(n!)^2} \sum_{j=0} \left[\frac{n!}{j!(n-j)!}\right]^2\right] \tag{S29}$$

$$= \Re\left[\log k - \sum_{n=1}^{\infty} \frac{1(2n-1)!}{k^{2n}(n!)^2} \sum_{j=0} \binom{n}{j}^2\right] \tag{S30}$$

$$= \Re\left[\log k - \sum_{n=1}^{\infty} \frac{1(2n-1)!}{k^{2n}(n!)^2} \sum_{j=0} \binom{n}{j}\binom{n}{n-j}\right] \tag{S31}$$

$$= \Re\left[\log k - \sum_{n=1}^{\infty} \frac{1(2n-1)!}{k^{2n}(n!)^2} \binom{2n}{n}\right] \tag{S32}$$

$$= \Re\left[\log k - \sum_{n=1}^{\infty} \frac{1(2n-1)!}{k^{2n}(n!)^2} \frac{(2n)!}{(n!)^2}\right] \tag{S33}$$

$$= \Re\left[\log k - \frac{2}{k^2} \sum_{n=0}^{\infty} \frac{1(2n+1)!(2n+2)!}{2k^{2n}((n+1)!)^4}\right] \tag{S34}$$

The ratio of the $n+1$-th term divided by the $n$-th term is

$$\frac{4(n+1)(2n+3)^2}{(n+2)^3} \cdot = \frac{16(n+1)(n+3/2)^2}{(n+2)^3} \cdot \tag{S35}$$

For a general $_4F_3$ function,

$$_4F_3 = \left[\begin{array}{c} a_1, a_2, a_3, a_4 \\ b_1, b_2, b_3 \end{array}; z\right] , \tag{S36}$$

the same quantity is

$$\frac{(a_1+n)(a_2+n)(a_3+n)(a_4+n)}{(n+1)(b_1+n)(b_2+n)(b_3+n)} \tag{S37}$$

The claim (8), then follows from the definition of $_pF_q$ functions and from a term-by-term comparison of the series expansion:

$$m(k) = \Re\left(\log k - \frac{2}{k^2} \, _4F_3\left[\begin{array}{c} 1,1,\frac{3}{2},\frac{3}{2} \\ 2,2,2 \end{array}; \frac{16}{k^2}\right]\right) .$$

## S3. INDEPENDENT CHECKS OF $t(k) = g(2-k)$

### A. Cross-check of Eq. (11) for $t(k)$ using two different expressions

As an exercise, we compare two different series expansions of $t(k)$ to low order as a cross-check of (11) of the main text. We have two distinct expressions for $t(k)$, namely Eq. (S17) in the appendix and the hypergeometric expression from the main text.

On the one hand, using (11) and (9) of the main text, we get from the hypergeometric expression to low order for positive $x$,

$$t(1/x) = \Re\left[\frac{1}{3}\log\left(-\frac{1}{x^3}\right) - 3x^2 + 4x^3 - \frac{45x^4}{2} + 72x^5 - 340x^6 + 1440x^7 + O\left(x^8\right)\right]. \tag{S38}$$

On the other hand (S17) gives us, for general $x \in \mathbb{C}$,

$$t(1/x) = \Re\left[\log\left(\frac{1}{x}\right) - 3x^2 + 4x^3 - \frac{45x^4}{2} + 72x^5 - 340x^6 + 1440x^7 + O\left(x^8\right)\right]. \tag{S39}$$

So the two expressions are in agreement to low order.

### B. Numerical verification of Eq. Eq. (11)

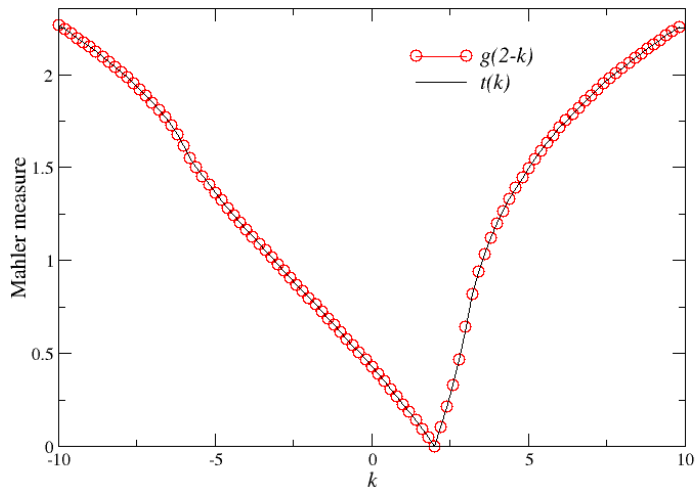

Figure S1: Plots of $t(k)$ and $g(2-k)$ for real $k \in \mathbb{R}$. The agreement of the curves is a reassurance that Eq. (11) of the main text is correct.


\clearpage


\begin{thebibliography}{mt1}


\bibitem{zudilin-book}
  F. Brunault F and W. Zudilin,
\href{https://doi.org/10.1017/9781108885553}
  {{\it  Many Variations of Mahler Measures: A
Lasting Symphony} (Cambridge University Press, Cambridge, 2020).}
%
%
%
%
%
  
  
\bibitem{smyth-book} J.  McKee and C. Smyth, \href
  {https://doi.org/10.1007/978-3-030-80031-4} {{\it Around the Unit
    Circle: Mahler Measure, Integer Matrices and Roots of Unity}
    (Springer, Cham, 2021).}


  
\bibitem{zahabi} A. Zahabi, Toric quiver asymptotics and Mahler
  measure: BPS states,
\href{https://doi.org/10.1007/JHEP07(2019)121}
{  J. High Energ. Phys. {\bf 2019}, 121
  (2019). }
%
%
%
%
%
%




\bibitem{stienstra-instanton} J. Stienstra, Mahler measure variations,
  Eisenstein series and instanton expansions, in Mirror symmetry V,
  AMS/IP Studies in Advanced Mathematics, vol. 38, eds. N. Yui,
  S.-T. Yau and J. D. Lewis (International Press \& American
  Mathematical Society, Providence, RI, 2006);
%
\href{https://arxiv.org/abs/math/0502193}  	{arXiv:math/0502193}
  
%

\bibitem{bao2022}
%
  %
J.  Bao, Y. H. He and A. Zahabi, A. Mahler Measure for a Quiver
Symphony,
\href{https://doi.org/10.1007/s00220-022-04404-y}
{Commun. Math. Phys. {\bf 394,}
573 
%
(2022).} 







  
  
\bibitem{chesi2013} G. Chesi,
    \href{https://doi.org/10.1109/ACC.2013.6580630} 
  {On the Mahler measure of matrix pencils, in
  {\it 2013 American Control Conference} (IEEE, Washington, 2013).}
%
%
%
%
%
%
%
%
%



\bibitem{guttmann2012}
A. J. Guttmann and M. D Rogers,
  Spanning tree generating functions and Mahler
  measures,
\href{doi:10.1088/1751-8113/45/49/494001}{  J. Phys. A: Math. Theor. {\bf 45,} 494001 
(2012).}



  
\bibitem{viswan2017-pre} 
  G. M. Viswanathan 
Correspondence between spanning trees and the Ising model on a square lattice,
\href{https://doi.org/10.1103/PhysRevE.95.062138}
{ Phys. Rev. E} {\bf 95},
  062138  (2017).





  
\bibitem{stienstra-dimer}
  J. Stienstra, 
Mahler measure, Eisenstein series and dimers,
    in Mirror symmetry V,
  AMS/IP Studies in Advanced Mathematics, vol. 38, eds. N. Yui,
  S.-T. Yau and J. D. Lewis (International Press \& American
  Mathematical Society, Providence, RI, 2006); 
\href{https://arxiv.org/abs/math/0502197}{arXiv:math/0502197} 

%
%
 %
%
%
%



\bibitem{wu-of-glasser} M. L.  Glasser and F. Y. Wu, On the Entropy of
  Spanning Trees on a Large Triangular Lattice,
\href{https://doi.org/10.1007/s11139-005-4847-9}
  {Ramanujan J.{\bf 10,}
  205–214 (2005).}
%

\bibitem{glasser-of-glasser}
  M. L. Glasser and G. Lamb, A lattice spanning-tree entropy function
\href{https://doi.org/10.1088/0305-4470/38/25/L02}
  {  J. Phys. A: Math. Gen. {\bf 38} L471 (2005).}





%
%
   
\bibitem{shrock-of-glasser}
%
R. Shrock and F. Y. Wu,
Spanning trees on graphs and lattices in d dimensions
\href{https://doi.org/10.1088/0305-4470/33/21/303}
{J. Phys. A: Math. Gen. 33 3881 
(2000).}



 \bibitem{glasser-cubic}
M. L. Glasser,
A note on a hyper-cubic Mahler measure and
associated Bessel integral
%
\href{https://doi.org/:10.1088/1751-8113/45/49/494002}
{J. Phys. A: Math. Theor. {\bf 45}, 494002 (2012).}



\bibitem{boyd1981} 
D. W. Boyd, Speculations concerning the range of Mahler’s measure,
\href{doi:10.4153/CMB-1981-069-5}{Canad. Math. Bull., {\bf 24}(4):
  453  (1981).}
%


  
  \bibitem{boyd1998} D. W. Boyd, Mahler’s measure and special values
    of L-functions,
\href{https://doi.org/10.1080/10586458.1998.10504357}{    Experiment. Math., {\bf 7}(1),
    37  (1998).}
%




    \bibitem{lalin2007}
M. N. Lalín and Mathew D. Rogers,
Functional equations for Mahler measures of genus-one curves 87–117
\href{http://dx.doi.org/10.2140/ant.2007.1.87}
{{Algebra and Number Theory \{\bf 1}(1),
87 (2007).}
%


      \bibitem{villegas-1999} 
F. R.  Villegas, Modular Mahler Measures I, in: S. D. Ahlgren,
G. E. Andrews, K. Ono (eds),
\href{https://doi.org/10.1007/978-1-4613-0305-3_2}
{{\it Topics in Number Theory. Mathematics
  and Its Applications}, vol 467 (Springer, Boston, 1999).}


\bibitem{lalin2013} 
  Matilde N. Lalín, Equations for Mahler measure and isogenies,
\href{https://doi.org/10.5802/jtnb.841}
 { J. Théor.
Nombres Bordeaux {\bf 25}(2), 387  (2013).}

\bibitem{ising-elliptic}
{A. Bostan, S. Boukraa, S. Hassani, M. van Hoeij, J.-M. Maillard, J.-A. Weil and N. Zenine},
%
  {The Ising model: from elliptic curves to modular forms and Calabi–Yau equations},
\href{https://dx.doi.org/10.1088/1751-8113/44/4/045204}
  {  {Journal of Physics A: Mathematical and Theoretical},
  {\bf 44}(4)  {045204} (2011).}



\bibitem
  {ising-l-function}
%
Y. M. Zinoviev,
Ising Model and $L$-Function,
\href{https://doi.org/10.1023/A:1005202114780}
{Theoretical and
Mathematical Physics {\bf 126,} 66
(2001).}

%


\bibitem{ising-sigma}
  %
  S. Boukraa, S. Hassani, J.-M, Maillard  and N. Zenine,
  From Holonomy of the Ising Model Form Factors to $n$-Fold Integrals and the Theory of Elliptic Curves,  
\href{https://doi.org/10.3842/SIGMA.2007.099}
     {  SIGMA {\bf 3} 099, 43 (2007).}
     %


%


\bibitem {viswan2021} 
G. M. Viswanathan,
The double hypergeometric series for the partition function of the 2D anisotropic Ising model,
\href{https://doi.org/10.1088/1742-5468/ac0f71}
{
J. Stat. Mech. {\ bf 2021}, 073104 (2021).}

     


\bibitem{viswan-entropy} G. M.  Viswanathan, M. A. G.  Portillo, E. P.
  Raposo, M. G. E.  da Luz,
%
What
Does It Take to Solve the 3D Ising Model? Minimal Necessary Conditions
for a Valid Solution.
\href{https://doi.org/10.3390/e24111665}
{Entropy {\bf 24,}
  1665.
(2022).}




\bibitem{baxter} 
R. J. Baxter,
\href{https://physics.anu.edu.au/research/ftp/_files/Exactly.pdf}    
    {{\it Exactly solved models in
  statistical mechanics} 
(Academic Press, London,1989).}




\bibitem{lehmer} 
D. H. 
Lehmer, Factorization of certain cyclotomic
functions,
\href
{doi:10.2307/1968172}
{Ann. Math. 2. {\bf 34}(3),
461
%
(1933).}
%
%


  

\bibitem{mahler1962} 
  K. Mahler, On some inequalities for polynomials in several variables,
\href{https://doi.org/10.1112/jlms/s1-37.1.341}
  {  J. London Math. Soc., {\bf 37}
%
  341 (1962).}





\bibitem{mahler-zeta}
  %
  B. J. Ringeling, Mahler measures, modular forms and hypergeometric
  functions,
  \href{https://hdl.handle.net/2066/297627}
       {Ph.D. thesis, Radboud University, Nijmegen (1993).}


  




  
\bibitem{g-singular}
A. Mellit,
Elliptic dilogarithms and parallel lines,
\href{https://doi.org/10.1016/j.jnt.2019.03.019}
{J. Number Theory {\bf 204}, 1 (2019);}
%
\href{https://doi.org/10.48550/arXiv.1207.4722}{arXiv:1207.4722 (2012).} 

\bibitem{m-singular}
Z. Tao, G. Xuejun and T. Wei,
{Mahler measures and $L$-values of elliptic curves over real quadratic fields},
  \href{https://arxiv.org/abs/2209.14717}{arXiv:2209.14717 (2022).}        





\bibitem{deninger} Deninger C. Deninger, Deligne periods of mixed
  motives, $K$-theory and the entropy of certain
  $Z^n$-actions.
\href{https://doi.org/10.1090/S0894-0347-97-00228-2}  {J. Amer. Math. Soc., 10(2):259–281, 1997.}








\bibitem{rogers2010} M. Rogers, Hypergeometric Formulas for
  Lattice Sums and Mahler Measures,
  \href{https://doi.org/10.1093/imrn/rnq240} {International
    Mathematics Research Notices, {\bf 2011}(17), 4027 (2011);}
\href{https://doi.org/10.48550/arXiv.0806.3590}
  { 	arXiv:0806.3590}





  
   \bibitem{hucht2011} A. Hucht, D. Grüneberg and F. M. Schmidt,
     Aspect-ratio dependence of thermodynamic Casimir forces,
     \href{https://doi.org/10.1103/PhysRevE.83.051101}{
       { Phys. Rev. E} {\bf 83},
  051101 (2011).}

  


%

%
%
%


\bibitem{onsager} L. Onsager, Crystal Statistics. I. A Two-Dimensional
  Model with an Order-Disorder Transition
  \href{https://link.aps.org/doi/10.1103/PhysRev.65.117}
       {Phys. Rev. {\bf 65}, 117 (1944).}



\bibitem{hexagon}  
R.M.F. Houtappel,
Order-disorder in hexagonal lattices,
\href{https://doi.org/10.1016/0031-8914(50)90130-3}
{Physica,
{\bf 16}(5), 425
(1950).}
%



\bibitem{poincare}
  H. Poincaré
  %
\href{https://archive.org/details/sciencemethod00poinuoft} {{\it Science
    and Method} (T. Nelson, London, 1914).}
%
%
%
%



\bibitem{yang2002}


  H.  Au-Yang and J. H. H. Perk,
%
  Correlation Functions and Susceptibility in the $Z$-Invariant Ising
  Model.
\href{https://doi.org/10.1007/978-1-4612-0087-1_2}
  {In: M. Kashiwara and T. Miwa, T. (eds), {\it MathPhys
    Odyssey 2001}, Progress in Mathematical Physics {\bf
  23} (Birkhäuser, Boston, 2002).}



  
\bibitem{perk1} J. H. H. Perk, Comment on Zhang, D. Exact Solution for
  Three-Dimensional Ising Model. Symmetry 2021, 13,
  1837,
  \href{https://doi.org/10.3390/sym15020374}
  {  Symmetry {\bf 15}(2), 374.
(2023).}



  
  
\bibitem{perk2}  
M. E. Fisher and J. H.H. Perk,
%
Comments concerning the Ising model and two letters by N.H. March,
%
\href{https://doi.org/10.1016/j.physleta.2015.09.055}
{Physics Letters A,
{\bf 380}(13)
%
1339
%
(2016).}



%
%
%

  
\bibitem{perk3}
%
J. H. H. Perk
%
Comment on 'Mathematical structure of the three-dimensional (3D) Ising model'
%
%
\href{https://dx.org/10.1088/1674-1056/22/8/080508}
{
Chinese Physics B {\bf 22}(8), 080508
(2013).}



  
\bibitem{perk4}
J. H. H. Perk,
%
Erroneous solution of three-dimensional (3D) simple orthorhombic Ising lattices
%
\href{https://journals.ltn.lodz.pl/Bulletin/issue/view/209/209}
{Bulletin de la Société des Sciences et des Lettres de Łódż, Série:
Recherches sur les Déformations
{\bf 62}(3), 45 (2013).}
%
%



  





%
  
  




\end{thebibliography}
\end{document}